\documentclass[a4paper,11pt]{article}
\pdfoutput=1 
\usepackage{jheppub} % for details on the use of the package, please
\usepackage[T1]{fontenc} % if needed
\usepackage{tabu}
\def\bp{\begin{pmatrix}}
\def\ep{\end{pmatrix}}
\def\ba{\begin{align}}
\def\ea{\end{align}}
\def\eps{\epsilon}

\def\l({\left(}
\def\r){\right)}
\def\be{\begin{equation}}
\def\ee{\end{equation}}
\newcommand{\VerySmall}{\fontsize{10}{11}\selectfont}

\definecolor{rust}{rgb}{0.8,0.2,0.2}
\definecolor{green}{rgb}{0.1,0.8,0.2}

\title{Comments on Hall transport from effective actions}

\author[]{Felix Haehl,}
\author[]{Mukund Rangamani}
\affiliation[]{Centre for Particle Theory \& Department of Mathematical Sciences,\\
                     Science Laboratories, South Road, Durham DH1 3LE, UK.}
\emailAdd{f.m.haehl@durham.ac.uk}
\emailAdd{mukund.rangamani@durham.ac.uk}

\abstract{We consider parity-odd transport in $2+1$ dimensional charged fluids restricting attention to the class of non-dissipative fluids. We show that there is a two parameter family of such non-dissipative fluids which can be derived from an effective action, in contradistinction with a four parameter family that can be derived from an entropy current analysis. The effective action approach  allows us to extract the adiabatic transport data, in particular the Hall viscosity and Hall conductivity amongst others, in terms of the thermodynamic functions that enter as `coupling constants'. Curiously, we find that Hall viscosity is forced to vanish, whilst the Hall conductivity is generically a non-vanishing function of thermodynamic data determined in terms of the hydrodynamic couplings.}

\begin{document}
\begin{flushright} \small{DCPT-13/19} \end{flushright}

\maketitle
\flushbottom

%~~~~~~~~~~~~~~~~~~~~~~~~~~~~~~~~~~~~~~~~~~~~~~~,
\section{Introduction}
\label{sec:intro}
%~~~~~~~~~~~~~~~~~~~~~~~~~~~~~~~~~~~~~~~~~~~~~~

Hydrodynamics, which is an effective description of long-wavelength, near-equilibrium physics, has received a renewed interest in recent years, owing to a concatenation of circumstances. The developments have partly been inspired by developments of the connections between gravity and hydrodynamics, a la the fluid/gravity correspondence \cite{Bhattacharyya:2008jc,Hubeny:2011hd} and by the potential for macroscopic signatures of parity violations and anomalies in transport phenomena \cite{Son:2009tf}. 

An interesting question that has garnered some attention is a definition of an autonomous theory of hydrodynamics viewed as an effective field theory. Traditional approaches to the subject have relied upon writing down conserved currents (for energy-momentum, charge etc.) as functionals of fluid dynamical variables (a normalized timelike $d$-velocity and local temperature, chemical potential) in a suitable gradient expansion. These constitutive relations expressing the currents are however not arbitrary. They are constrained by demanding that the local form of the second law of thermodynamics be valid for arbitrary fluid flows. This philosophy has been appreciated for many decades now \cite{landau}, and has proven to be immensely successful in delineating the constraints on constitutive relations in a wide variety of cases. 

An interesting upshot of these developments is the appreciation of the fact that we have two distinct types of transport coefficients: the first kind are the genuine hydrodynamic transport coefficients (eg., viscosity and conductivity), wheres the second kind correspond to thermodynamic response parameters. The distinction between the two is that the former are indeterminable in equilibrium, whereas the latter can, simply by suitably placing the fluid in stationary (time independent) backgrounds. Consequentially the latter are adiabatic, they do not contribute to entropy production in hydrodynamics; as a result they can be encapsulated in a partition function, just as usual thermodynamic parameters \cite{Banerjee:2012iz, Jensen:2012jh}.  

However, it is far from clear that these constraints are sufficient. In part, the opacity in formulating a clean autonomous theory of hydrodynamics is traceable to the fact that the second law of thermodynamics is mysterious from a microscopic perspective. It is therefore desirable to have at least a set of toy models wherein one might try to explore hydrodynamic constraints from a more conventional effective action perspective. A natural candidate which allows such explorations is the theory of {\em non-dissipative hydrodynamics}. 

The basic framework for such a theory for a long time has been the ideal fluid which is the prototypical example of a non-dissipative fluid (see e.g., \cite{Jackiw:2004nm} for a review). Its utility in understanding hydrodynamics from a modern perspective was brought to fore in \cite{Dubovsky:2011sj}.\footnote{See \cite{Dubovsky:2005xd} for earlier work, \cite{Endlich:2012vt,Grozdanov:2013dba} for attempts to include dissipation and \cite{Andrianopoli:2013dya,Andrianopoli:2013vba} for supersymmetric extensions.} It was furthermore shown to allow access to understanding anomalous hydrodynamical transport (in 2 spacetime dimensions) in \cite{Dubovsky:2011sk}. Building on these works in \cite{Bhattacharya:2012zx} the constraints on neutral fluids were examined in some detail and it was argued that the effective action for non-linear non-dissipative fluids constrains transport more strongly than the traditional approach employing the entropy current. Note that since we are concerned with non-dissipative fluids, we require that the hydrodynamical system accommodate an exactly conserved entropy current;\footnote{The constraints on generic neutral fluids arising from requiring the local form of second law to hold have been derived to second order in the gradient expansion in \cite{Bhattacharyya:2012nq}.} in short all flows of the system produce no entropy.

A natural extension of the above program of analyzing constraints on non-dissipative fluids is to include additional conserved charges, say a conserved $U(1)$ charge. While the analysis of parity-even charged fluids is possible, here we examine a simpler system of parity-odd charged fluids in $2+1$ dimensions motivated by physical interest in 
Hall transport. Indeed, one of the first applications of the effective action formalism was to understand Hall viscosity in  \cite{Nicolis:2011ey} (see also \cite{Bhattacharya:2012zx}). We take a critical look at this particular transport phenomenon and comment on Hall conductivity through our analysis.

As we mention part of our motivation is to contrast the effective action approach against an entropy analysis. Fortunately for us \cite{Jensen:2011xb} analyze quite thoroughly the constraints of the second law on parity-odd transport in the case of interest and by suitably adapting their computation to non-dissipative systems, we will be able to compare the two approaches. As in \cite{Bhattacharya:2012zx} we will find that the effective action is more constraining than the existence of a conserved entropy current. The effective action turns out to be parameterized by two thermodynamic functions which determine the Hall conductivity and viscosity. Surprisingly, we find the Hall viscosity vanishes. In contrast the entropy current analysis is insensitive to these transport coefficients. We find the vanishing of Hall viscosity puzzling and offer some speculations regarding the effective action approach to non-dissipative hydrodynamics.

The outline of this paper is as follows: in \S\ref{sec:nondis} we quickly remind the reader of the basic ingredients of the effective action formalism for non-dissipative fluids. We then analyze parity-odd charged fluids in $2+1$ dimensions using this formalism at the first non-trivial order in the gradient expansion in \S\ref{sec:charged3d} and contrast the result with a direct entropy current analysis (by adapting the results of \cite{Jensen:2011xb}). We conclude with a discussion of our results and some observations regarding Hall transport in \S\ref{sec:discuss}.

%~~~~~~~~~~~~~~~~~~~~~~~~~~~~~~~~~~~~~~~~~~~~~~~
\section{A brief review of non-dissipative fluids}
\label{sec:nondis}
%~~~~~~~~~~~~~~~~~~~~~~~~~~~~~~~~~~~~~~~~~~~~~~

We begin with a lightning review of the basic ingredients of the effective action approach. Following \cite{Dubovsky:2011sj,Bhattacharya:2012zx} degrees of freedom for a charged fluid in $d$ dimensions are $d-1$ fundamental fields $\{\phi^I\}$ which give the position of local fluid elements in physical space (coordinates $x^\alpha$) and an additional phase label $\psi$ which characterizes the charge contained in the fluid element. The fields $\{\phi^I,\psi\}$ are Lorentz scalars and enjoy reparametrization invariance
\begin{align}
\phi^I \to \xi^I(\phi) \,,\;\; {\rm with}\;\; \text{Jacobian}(\xi,\phi) = 1 \,,\qquad \psi \longrightarrow \psi + \mathfrak{f}(\phi^I) \, .
\label{sdiffphipsi}
\end{align}	
The latter symmetry, which has been called {\em chemical shift invariance}, generalizes the translational symmetry $\psi \to \psi +c$ for a phase field, to a local field dependent translation, while the former asserts that the volume of the $\phi^I$ configuration manifold is unchanged. Following \cite{Bhattacharya:2012zx} we will refer to these symmetries as $\widetilde{\text{Sdiff}({\cal M}_{\phi,\psi})}$, to indicate that we are concerned with volume preserving diffeomorphisms supplemented with field translations in the configuration manifold ${\cal M}_{\phi,\psi}$. The fields $\{\phi^I,\psi\}$ will appear derivatively coupled in our effective action, signifying that they should be viewed as Goldstone modes for the embedding of the fluid in the background spacetime; thus $[d\phi^I] = [d\psi] =0 $ is the canonical assignment of scaling dimensions for these field. 

There are two important consequences of the symmetry $\widetilde{\text{Sdiff}({\cal M}_{\phi,\psi})}$. Firstly, it guarantees that that dynamical equations of motion are isomorphic to the conservation of the energy-momentum and charge currents. Secondly, there is a kinematically conserved current which is identified with the entropy current of the fluid dynamics:
\begin{align}
 J_\text{S}^\beta = \frac{1}{(d-1)!}\; \eps^{\beta \alpha_1 ... \alpha_{d-1}}\; \eps_{I_1...I_{d-1}} \; \prod_{j=1}^{d-1}
    \partial_{\alpha_j} \phi^{I_j} \,, \qquad \nabla_\alpha J_\text{S}^\alpha = 0 .
\end{align}
The basic fluid dynamical parameters for a neutral fluid are constructed out of this invariant $J_\text{S}^\alpha$. The entropy density $s$ and the fluid dynamical velocity field $u^\alpha$ and the chemical potential 
$\mu$ that couples to the conserved $U(1)$ charge are simply defined as  
\begin{align}
J_\text{S}^\alpha = s\, u^\alpha \, , \qquad s = \sqrt{-\, g_{\alpha\beta}\, J_\text{S}^\alpha\, J_\text{S}^\beta}  \, , \qquad\mu = u^\alpha\, D_\alpha \psi \,,
\end{align}
In the expression for the chemical potential we allow for a non-dynamical background gauge field ${\bf A} = A_\alpha\, dx^\alpha$ by using the background covariant derivative ${\bf D} \psi = d \psi + {\bf A}$. These objects are all invariant under $\widetilde{\text{Sdiff}({\cal M}_{\phi,\psi})}$.\footnote{Gauge transformations for our system are simply $\psi \to \psi - \lambda(x)$ together with ${\bf A} \to {\bf A} + d\lambda$. To see the invariance of the chemical potential under the chemical shift transformation, it is useful to note that  $J^\alpha_\text{S}$ is co-moving with the fluid elements i.e., $J_\text{S}^\alpha \nabla_\alpha \phi^I=0$.}

To see that our identification of the thermodynamics is correct, we start with the most general action invariant under $\widetilde{\text{Sdiff}({\cal M}_{\phi,\psi})}$ to leading order. This is simply \cite{Dubovsky:2011sj}:
\begin{align}
 S_0 = \int d^3x\,\sqrt{-g}\,  f(s,\mu) \, .
\label{S0def}
\end{align}
Abbreviating $f_{,s}\equiv \tfrac{\partial}{\partial s} f$ and $f_{,\mu} \equiv \tfrac{\partial}{\partial \mu}f$, we find the following stress tensor and charge current\footnote{These are defined by varying the background sources. To wit, $T^{\alpha\beta} = \frac{2}{\sqrt{-g}} \frac{\delta S}{\delta g_{\alpha\beta}}$ and $J^\alpha = \frac{1}{\sqrt{-g}} \frac{\delta S}{\delta A_\alpha}$.}
\begin{align}
T^{\alpha\beta}_{(0)} &= 
                  \left( f-s  f_{,s} \right) g^{\alpha\beta} + \left(
                    \mu\, f_{,\mu}- s \,f_{,s}\right) u^\alpha u^\beta  
                 \equiv \varepsilon \, u^\alpha u^\beta + P \, P^{\alpha\beta} \, ,\label{ConRel01} \\
J^\alpha_{(0)} &= f_{,\mu}\, u^\alpha \equiv q\, u^\alpha \, , \label{ConRel02}
\end{align}
where we identified the energy density $\varepsilon =(\mu\,  f_{,\mu}-f)$, the pressure $P=(f-s f_{,s})$ and the charge  density $q=f_{,\mu}$. 
The projector orthogonal to $u^\alpha$, $P^{\alpha\beta} \equiv g^{\alpha\beta} +u^\alpha u^\beta$, is introduced for later convenience.
Eqs.\ (\ref{ConRel01}, \ref{ConRel02}) are just the constitutive relations for a charged perfect fluid.

%~~~~~~~~~~~~~~~~~~~~~~~~~~~~~~~~~~~~~~~~~~~~~~~
\subsection{Absence of partity-even non-dissipative terms at first order}
\label{sec:foeven}
%~~~~~~~~~~~~~~~~~~~~~~~~~~~~~~~~~~~~~~~~~~~~~~ 
 
 We now illustrate the deviations away from an ideal fluid systematically in a low energy gradient expansion in a somewhat trivial example of partity-even charged fluids. In the hydrodynamic expansion, the ideal fluids occur at the zeroth order in gradients. In general we write
 \begin{equation}
T^{\alpha\beta} = T^{\alpha\beta}_{(0)} + \Pi^{\alpha\beta} \,,\qquad J^\alpha  = J^\alpha_{(0)} + \nu^\alpha
 \label{}
 \end{equation}	
 signifying the split between the ideal and higher order contributions to the hydrodynamic constitutive relations.\footnote{We impose no a priori frame condition on fluid dynamics, since the effective action analysis is adapted to the entropy frame whereas the entropy current analysis is simple in the Landau frame.}
 
 One might wonder if it is possible to correct the fluid constitutive relations, staying on the locus of non-dissipative fluids at first order. It is intuitively clear that this cannot be achieved physically, since the first order transport coefficients corresponds to the viscosities (shear and bulk) and conductivity (electro-thermal). Both of these explicitly rely on the fluid being dissipative and vanish for non-dissipative fluids.
This can be immediately checked in the effective action framework. There are two non-trivial scalars that one can construct out of the fields $\{\phi^I,\psi\}$ respecting the symmetries,
\begin{align}
 J^\alpha_\text{S}\, \nabla_\alpha \mu  \,,\qquad 
 J^\alpha_\text{S} \nabla_\alpha s \,. \label{Listing}
\end{align}
That these are the only first order parity-even scalars at our disposal is clear from the fact that the $s$ and $\mu$ are invariant under $\widetilde{\text{Sdiff}({\cal M}_{\phi,\psi})}$ and $\nabla_\alpha J^\alpha_\text{S} =0$.\footnote{Chemical shift invariance \eqref{sdiffphipsi} precludes us from using the transverse vector $P^{\alpha\beta}\nabla_\beta \psi$.}
However, using the zeroth order equations of motion, one can
show that 
\begin{align}
 J^\alpha_\text{S} \,\nabla_\alpha \mu = \frac{1}{s} \left[ \left( \frac{\partial P}{\partial \rho} \right)_\varepsilon +  
   \mu \left( \frac{\partial P}{\partial \varepsilon} \right)_\rho \right] J^\alpha_\text{S} \nabla_\alpha s \, .\label{Identity0}
\end{align} 
As a result the most general parity-invariant first order action that one could write is
\begin{align}
 S_1^\text{(even)} &= \int d^3x\, \sqrt{-g} \, f_1(s,\mu) \;J^\alpha_S \nabla_\alpha s  \,. \label{S1}
\end{align}

If $f_1(s,\mu)=f_1(s)$ is independent of the chemical potential, then the same argument as given in \cite{Bhattacharya:2012zx} for uncharged fluids shows that the integrand can be written as a total derivative and therefore be ignored (since it doesn't affect the equations of motion).

However, if $f_1(s,\mu)$ does depend on $\mu$, we need to go a bit further in order to show that the physical consequences 
of $S_1^\text{(even)}$ are actually trivial and we can thus disregard $S_1^\text{(even)}$. 
The contributions of $S_1^\text{(even)}$ to the constitutive relations are easily determined to be 
\begin{align}
 \Pi^{\alpha\beta}_{\text{(1,even)}} &=  s \,f_{1,\mu} \,J^\gamma_\text{S}\, \nabla_\gamma \mu \;  P^{\alpha\beta}
      - s^2 \,\mu \,f_{1,\mu}\, \Theta \, u^\alpha u^\beta \,, \label{FakeT}\\
 \nu^\alpha_\text{(1,even)} &= -s^2\, f_{1,\mu} \,\Theta \, u^\alpha \,, \label{FakeJ}
\end{align}
where we used that $J^\alpha_\text{S}\, \nabla_\alpha s = - s^2 \Theta$ (see Table \ref{table:3Dclassification} for definitions of various fluid dynamical tensors). 
These contributions do not lead to any transport and are just an artifact of the choice of a frame. One can note this since all the terms in \eqref{FakeT} and \eqref{FakeJ} do is shift the zeroth order thermodynamic functions $\{\varepsilon, P, \rho \}$.  More formally,  using (\ref{Identity0}), all the frame invariant information (cf., \S\ref{sec:3compare}) contained in the expressions 
(\ref{FakeT}, \ref{FakeJ}) can easily be shown to vanish. Since we will be interested in non-trivial transport phenomena, we will not need to take into account the spurious effects of $S_1^\text{(even)}$
in the following.\footnote{A simpler argument which establishes the result is to note that by allowing an appropriate amount of $J^\alpha_\text{S}\,\nabla_\alpha \mu$ we can covert \eqref{S1} into a total derivative.} This shows that for the following first order analysis, we can ignore any terms which preserve parity.

%~~~~~~~~~~~~~~~~~~~~~~~~~~~~~~~~~~~~~~~~~~~~~~~~~~~~~~~~~~
\section{Charged parity-odd $2+1$ dimensional fluids}
\label{sec:charged3d}
%~~~~~~~~~~~~~~~~~~~~~~~~~~~~~~~~~~~~~~~~~~~~~~~~~~~~~~~~~~

Having described the basic structure in constructing effective actions for non-dissipative hydrodynamics, we now turn to an analysis for the specific case of $2+1$ dimensional charged fluids. Since there is no non-trivial transport in the parity-even section as established in \S\ref{sec:foeven}, we shall focus on parity-odd fluids. In three dimensions these contributions are the leading corrections to the ideal fluid, which we will first establish in \S\ref{sec:3eaodd}. Following this effective action analysis we will address the question of non-dissipative transport using the entropy current formalism in \S\ref{sec:3ecodd}  and contrast the two approaches in \S\ref{sec:3compare}.

%~~~~~~~~~~~~~~~~~~~~~~~~~~~~~~~~~~~~~~~~~~~~~~~~~~~~~~~~~~
\subsection{Effective action for parity-odd transport}
\label{sec:3eaodd}
%~~~~~~~~~~~~~~~~~~~~~~~~~~~~~~~~~~~~~~~~~~~~~~~~~~~~~~~~~~

At first order we have the following differential forms from which we can construct terms in the effective action: one-forms ${\bf u}, \; {\bf A},  \; ds, \; d\mu$ and the two-forms $d{\bf u} , \;  d{\bf A}$. We need to use these data to construct one-derivative three-forms which are both invariant under the shift symmetry \eqref{sdiffphipsi} and under gauge transformations. The only such terms are ${\bf u} \wedge d{\bf u}$ and ${\bf u}\wedge d{\bf A}$. 

Therefore, the most general first order action is parameterized by two functions ${\mathfrak w}(s,\mu)$ and ${\mathfrak b}(s,\mu)$:
\begin{align}
  S_1 = \int d^3x\, \sqrt{-g} \, \left[ {\mathfrak w}(s,\mu) \, \eps^{\rho\sigma\lambda}\, u_\rho \nabla_\sigma u_\lambda
        +{\mathfrak b}(s,\mu) \, \eps^{\rho\sigma\lambda} \,u_\rho \nabla_\sigma A_\lambda \right] \, . \label{1stOrderAction}
\end{align}
Varying this action with respect to the sources, we obtain the parity-odd contributions to the energy-momentum tensor and current:\footnote{Variations 
with respect to $g_{\alpha\beta}$ are given by $\delta u^\mu = \tfrac{1}{2} u^\mu \, u^\alpha u^\beta \delta g_{\alpha\beta}$ 
and $\delta A_\mu = 0$.}
\begin{align}
\Pi^{\alpha\beta}_{(1)}= &\; -[s \, {\mathfrak w}_{,s}\, \Omega +s\, {\mathfrak b}_{,s} \, B]  P^{\alpha\beta} + \left[(2\, {\mathfrak w} + \mu\,  
{\mathfrak w}_{,\mu})\, \Omega+  ({\mathfrak b} + \mu \, {\mathfrak b}_{,\mu})\, B\right]  u^\alpha u^\beta
             \notag \\         
         & + 4\, {\mathfrak w} \,\eps^{(\alpha\rho\sigma} u^{\beta)} \nabla_\rho u_\sigma 
         + 2 \,{\mathfrak b} \,\eps^{(\alpha\rho\sigma} u^{\beta)} \,\nabla_\rho A_\sigma 
         -  2\,\eps^{(\alpha\rho\sigma} \,u_\rho u^{\beta)} ({\mathfrak w}_{,s} \nabla_\sigma s +{\mathfrak w}_{,\mu} 
            \nabla_\sigma \mu )  \, , \label{Pi1} \\
\nu^\alpha_{(1)} = & \; 
         {\mathfrak b}\,\eps^{\alpha\rho\sigma} \nabla_\rho  u_\sigma -\eps^{\alpha\rho\sigma} u_\rho ({\mathfrak b}_{,s} \,\nabla_\sigma s +{\mathfrak b}_{,\mu}\, \nabla_\sigma \mu)
          + \left( {\mathfrak w}_{,\mu}\, \Omega + {\mathfrak b}_{,\mu} \, B\right)u^\alpha \, ,
         \label{Current1}
\end{align}
where the parity-odd scalars $\Omega$ and $B$ are defined in Table \ref{table:3Dclassification}.
We remind the reader that since $J_S^\alpha = s\, u^\alpha$ for the theory under consideration, it follows that the above results are in the entropy frame.

The analysis above generalizes the discussion of neutral parity-odd fluids. These were originally described in \cite{Nicolis:2011ey} and revisited briefly in \cite{Bhattacharya:2012zx}. These results are imminently recovered by setting ${\mathfrak b}(s,\mu) =0$ and replacing ${\mathfrak w}(s,\mu) \to {\mathfrak w}(s)$.

%~~~~~~~~~~~~~~~~~~~~~~~~~~~~~~~~~~~~~~~~~~~~~~~~~~~~~~~~~~
\subsection{Constraints from entropy current analysis}
\label{sec:3ecodd}
%~~~~~~~~~~~~~~~~~~~~~~~~~~~~~~~~~~~~~~~~~~~~~~~~~~~~~~~~~~

We will now study the constraints on $2+1$ dimensional partity-odd fluids which follow from just the vanishing divergence of the most general entropy current. 
This calculation has been carried out in \cite{Jensen:2011xb} and we will 
closely follow their analysis (for the same analysis in equilibrium, see \cite{Banerjee:2012iz}). The main difference in our analysis is that we will demand that the entropy current be divergence free, in contradistinction to the usual story described in \cite{Jensen:2011xb} where it is only required to have non-negative divergence.

% {\renewcommand{\arraystretch}{1.4}
\begin{table}
\VerySmall{
\centering
{\tabulinesep=1.5mm
\begin{tabu}{|c||c|c|c|}	
\hline
 & Data & Eqs.\ of motion & On-shell independent
     \\ 
    \hline \hline
Scalars & $\Theta \equiv \nabla_\alpha u^\alpha$ & $\nabla_\mu J^\mu = 0$ & 
 \\ 
        & $u^\alpha\nabla_\alpha T$ & $u_\mu \nabla_\nu T^{\mu\nu} = u_\mu  F^{\mu\nu} J_\nu$ & $\Theta$ \\
        & $u^\alpha\nabla_\alpha \mu$ & & \\
    \hline
Pseudo- & $\Omega \equiv \eps^{\mu\nu\rho}u_\mu\nabla_\nu u_\rho$ & & $\Omega$ \\ 
Scalars & $B \equiv \tfrac{1}{2} \eps^{\mu\nu\rho} u_\mu F_{\nu\rho}$ & & $B$  \\
    \hline
Vectors & ${\mathfrak a}^\mu=u^\alpha\nabla_\alpha u^\mu$ & &  
          $U_1^\mu \equiv {\mathfrak a}^\mu$ \\
        & $E^\mu \equiv F^{\mu\nu}u_\nu$ & $P_{\alpha\nu}\nabla_\mu T^{\mu\nu} = P_{\alpha\nu} F^{\nu\mu} J_\mu$ &  $U_2^\mu\equiv E^\mu$ \\
        & $P^{\mu\nu} \nabla_\nu T$ &  & $U_3^\mu\equiv P^{\mu\nu}\nabla_\nu (\tfrac{\mu}{T})- \tfrac{E^\mu}{T}$ \\
        & $P^{\mu\nu}\nabla_\nu (\tfrac{\mu}{T})$ & &  \\
     \hline
Pseudo- & $\Sigma^{\mu\nu}\nabla_\mu T$ & 
           & $\tilde V_1^\mu
            \equiv \Sigma^{\mu\nu}\nabla_\nu T$ \\
Vectors & $\Sigma^{\mu\nu} U_{1\nu}$ & 
$\Sigma_{\alpha\nu}\nabla_\mu T^{\mu\nu} = \Sigma_{\alpha\nu} F^{\nu\mu} J_\mu$
& $\tilde V_2^\mu\equiv \Sigma^{\mu\nu} U_{2\nu}$ \\
        & $\Sigma^{\mu\nu} U_{2\nu}$ & & $\tilde V_3^\mu \equiv \Sigma^{\mu}_\nu\left( U_{3}^\nu + \tfrac{1}{T} U_{2}^\nu\right)$ \\
        & $\Sigma^{\mu\nu} U_{3\nu}$ & & \\
     \hline
Tensors & $\sigma^{\mu\nu} \equiv P^{\mu\alpha} P^{\nu\beta} \left(\nabla_{(\alpha} u_{\beta)}-\frac{1}{2}\, \Theta \,
          P_{\alpha\beta}\right)$ & & $\sigma^{\mu\nu}$ \\
     \hline
Pseudo- & $\tilde \sigma^{\mu\nu} = \eps^{\alpha\rho(\mu} u_\alpha \sigma_\rho{}^{\nu)}$ & & 
          $\tilde \sigma^{\mu\nu}$ \\
Tensors & & & \\
\hline
\end{tabu}}
}
\caption{The right column contains a complete on-shell basis of $2+1$ dimensional fluid and background (gauge field) data at first order in derivatives.
They are obtained from off-shell independent data (left column) by using equations of motion (middle column).
We use the parity-odd projector $\Sigma^{\mu\nu}\equiv \eps^{\mu\rho\nu}u_\rho$. We also  abbreviate ${\bar \mu} = \mu/T$ for brevity.}
\label{table:3Dclassification}
\end{table}

We start be writing the most general entropy current allowed by symmetries in the basis of first order fluid data which is presented in Table \ref{table:3Dclassification}. 
In Landau frame this current takes the form\footnote{Note that we use a different sign convention than \cite{Jensen:2011xb} for some of the quantities defined in Table \ref{table:3Dclassification}.}
\begin{align}
 J_\text{S}^\alpha &
     = s\, u^\alpha- \frac{1}{T}\,  u_\beta \, \Pi^{\alpha\beta}
    -\frac{\mu}{T} \, \nu^\alpha + J_{\text{S},\,(1)}^\alpha \label{SCurrent}
\end{align}
with the dissipative parts in transport parameterized at first order as
\begin{align}
\Pi^{\alpha\beta}_{(1)} &= (-\zeta \, \Theta +\tilde \chi_B \, B +\tilde \chi_\Omega \, 
     \Omega)\; P^{\alpha\beta} - \eta \, \sigma^{\alpha\beta} - \tilde \eta \; \tilde \sigma^{\alpha\beta} \; ,\label{TEntr} \\
\nu^\alpha_{(1)} &= -\chi_T \; T \,U_1^\alpha + \chi_E \; U_2^\alpha -
      \sigma \; T \, U_3^\alpha + \tilde \chi_T \; \tilde V_1^\alpha+ \tilde \chi_E\; \tilde V_2^\alpha +
      \tilde \sigma \, (\tilde V_2^\alpha-T \,\tilde V^\alpha_3)   \,. \label{JEntr}
\end{align}
Note that in addition to the standard coefficients: conductivity ($\sigma$), shear and bulk viscosities ($\eta$, $\zeta$)  we have the parity-odd transport coefficients introduced in \cite{Jensen:2011xb}.

The general first order correction to the entropy current including parity-odd terms takes the form:
\begin{align}
 J^\alpha_{\text{S}, \,(1)} = \nu_0 \,\Theta\, u^\alpha + \sum_{i=1}^3 \; \nu_i\, U_i^\alpha + \sum_{i=1}^5 \; \tilde \nu_i\,
    \tilde V_i^\alpha \; ,
	\label{poddec}
\end{align}
where  we introduce
\begin{align}
  \tilde V_4^\alpha \equiv  -B\, u^\alpha + \Sigma^{\alpha\beta}\, U_{2\beta} \quad \text{ and } \quad 
  \tilde V_5^\alpha \equiv -\Omega \, u^\alpha-\Sigma^{\alpha\beta}\, U_{1\beta}\; .
\end{align}
While we a priori have $9$ independent parameters $\{\nu_i,\tilde \nu_i\}$ (which are functions of $T,\mu$) in \eqref{poddec}, we note that one of them is redundant. The pseudo-vector $\eps^{\alpha\rho\sigma} \nabla_\sigma (\tilde \nu_5\, u_\rho)$ is divergence free and can be expressed in terms of the others as \cite{Jensen:2011xb}
\begin{align}
\eps^{\alpha\rho\sigma} \nabla_\sigma (\tilde \nu_5\,  u_\rho)
 = \partial_T \tilde\nu_5 \, \tilde V_1^\alpha + \partial_{\bar\mu} \tilde\nu_5
   \, \tilde V_3^\alpha - \tilde \nu_5 \tilde V_5^\alpha \, , \label{DivFreeVector}
\end{align}
where $\bar\mu \equiv \mu/T$. Being divergence free, it can be added to $J_{\text{S},\,(1)}^\alpha$  leaving unchanged the requirement $\nabla_\alpha J_\text{S}^\alpha = 0$. This has the effect of removing the explicit $\tilde\nu_5$-term and shifting $\tilde \nu_1 \rightarrow \tilde\nu_1^* \equiv \tilde \nu_1+ \partial_T \tilde \nu_5$ and $\tilde \nu_3 \rightarrow \tilde\nu_3^* \equiv \tilde \nu_3+ \partial_{\bar\mu} \tilde \nu_5$. 

We can now calculate the divergence of the entropy current \eqref{poddec} and  express it as a linear combination of terms of second order gradient data. We start by writing down the pieces which contain independent genuinely second order fluid data (i.e., second order tensors which are not products of first order tensors). Such contributions arise solely from $J^\alpha_{S,(1)}$ and are given as
\begin{align}
\nabla_\alpha J^\alpha_\text{S}|_{\text{2-}\partial}
     &=\nabla_\alpha J^\alpha_{\text{S},\,(1)} \notag \\&= \left(\nu_2-\frac{\nu_3}{T}\right) \nabla_\alpha E^\alpha + \nu_3  \,
     P^{\alpha\beta} \nabla_\alpha \nabla_\beta \,\bar\mu +(\nu_0+\nu_1)\,  u^\alpha\nabla_\alpha \Theta
  \notag \\
&\qquad\qquad +\;\tilde \nu_2\; u^\alpha \nabla_\alpha B-\nu_1 \; u^\alpha u^\beta R_{\alpha\beta}  \, ,
\end{align}
where $R_{\alpha\beta}$ is the Ricci tensor of the background. All these terms are independent of each other and also 
independent of the rest of the divergence of the entropy current which are 
terms of products of first order pieces. Therefore, we conclude that\footnote{Note that these terms will have to vanish even for the weaker requirement that we have non-negative divergence of the entropy current.} 
\begin{align}
\tilde \nu_2 = \nu_0 = \nu_1 = \nu_2 = \nu_3 = 0 \, .
\end{align} 
The remaining part of the divergence of the entropy current can be computed to be
\begin{align}
   \nabla_\alpha J^\alpha_\text{S} &=  -\left[\frac{1}{T} \tilde \chi_\Omega
        - T \frac{\partial P}{\partial \varepsilon} \tilde \nu_1^*
        - \frac{1}{T} \frac{\partial P}{\partial q} \tilde \nu_3^* \right] \Theta\, \Omega
      - \left[ \frac{1}{T} \tilde \chi_B - T \frac{\partial P}{\partial \varepsilon} 
        \partial_T \tilde \nu_4 - \frac{1}{T} \frac{\partial P}{\partial q} \partial_{\bar\mu}
        \tilde \nu_4 \right] \Theta\, B \notag \\
   &\quad + \left[ \frac{qT}{\varepsilon+P} (\partial_T \tilde \nu_3^*-\partial_{\bar\mu}\tilde \nu_1^*)
        -\partial_{\bar\mu} \tilde\nu_4 + \frac{qT^2}{\varepsilon+P} \partial_T \tilde\nu_4
        +\tilde \chi_E \right] \tilde V_{2\alpha} U_3^\alpha \notag \\
   &\quad -\left[ \frac{qT}{\varepsilon+P} \tilde\nu_1^* - \frac{\tilde\nu_3^*}{T} - (\partial_{\bar\mu}
        \tilde\nu_1^* -\partial_T \tilde\nu_3^*)+ \tilde \chi_T\right] \tilde V_{1\alpha} U_3^\alpha   \notag \\
   &\quad + \left[ \frac{\tilde\nu_3^*}{T} + \partial_{\bar\mu} \tilde \nu_1^* - \partial_T
        \tilde \nu_3^* - T \partial_T \tilde\nu_4 \right] \tilde V_{2\alpha} U_1^{\alpha}  \notag \\
   &\quad -\chi_E \, U_2^\alpha U_{3\alpha} +T\chi_T U_1^\alpha U_{3\alpha} + \left(\sigma+\frac{qT}{\varepsilon+P}\chi_T\right) T \; U_3^\alpha U_{3\alpha} 
        +\frac{\eta}{2T} \; \sigma^{\alpha\beta} \sigma_{\alpha\beta}+\frac{\zeta}{T} \Theta^2\; .
\end{align}

To ensure that we have a divergence-free entropy current, we require that all the coefficients in front of the independent second order scalars vanish. We then find 
a system of equations, which can be solved explicitly. Following \cite{Jensen:2011xb} it is convenient to parameterize the result in terms of two  functions
${\mathfrak M}_B(T,\bar\mu)$, ${ \mathfrak M}_\Omega(T,\bar\mu)$.\footnote{In \cite{Jensen:2011xb} the solution is paramterized by three functions; in addition to the two we use here they introduce ${\mathfrak F}_\Omega(T)$. However, this can be absorbed into ${\mathfrak M}_\Omega$ and does not contain independent information.}
We first realize that the transport coefficients in the last line should all vanish independently. Then the coefficient of ${\tilde V}_{2\alpha} U_1^\alpha$ allows a single relation between entropy parameters $\{ {\tilde \nu}_3^*, {\tilde\nu}_1^*, {\tilde \nu}_4\}$
which we solve in terms of auxiliary functions $\{{\mathfrak M}_B , {\mathfrak M}_\Omega\}$. The remaining equations then simply determine $\{{\tilde \chi}_T, {\tilde \chi}_E, {\tilde \chi}_\Omega, {\tilde \chi}_B\}$ in terms of the functions we introduced. In short
\begin{align}
  \tilde\nu_4 &= \frac{1}{T} {\mathfrak M}_B \, , \qquad \tilde\nu_3^* = \frac{1}{T}\, 
      \partial_{\bar\mu} {\mathfrak M}_\Omega - {\mathfrak M}_B \, , \qquad
      \tilde \nu_1^* = \frac{1}{T} \partial_T {\mathfrak M}_\Omega - \frac{2}{T^2}
      {\mathfrak M}_\Omega , \notag \\
  \tilde \chi_T &=  T\, \partial_T \tilde\nu_4 - \frac{q\,T}{\varepsilon+P} \tilde\nu_1^* \, , \qquad
  \tilde \chi_E = \partial_{\bar\mu} \tilde \nu_4 - \frac{q}{\varepsilon+P} \tilde\nu_3^* \, , \notag \\
  \tilde \chi_\Omega &= T^2\,\frac{\partial P}{\partial \varepsilon} \, \tilde\nu_1^* + \frac{\partial P}{\partial
      q}  \tilde \nu_3^* \, , \qquad
  \tilde \chi_B = T^2 \,\frac{\partial P}{\partial \varepsilon} \,\partial_T \tilde\nu_4
      + \frac{\partial P}{\partial \rho} \partial_{\bar\mu} \tilde \nu_4 \, , \notag \\
   \chi_E &= \chi_T =\sigma = \eta = \zeta =  0 \, ,  \label{System1}
\end{align}
The transport coefficients $\tilde \sigma$ and $\tilde \eta$ remain unconstrained by the requirement of a vanishing divergence of the entropy current. They simply dropped out of the entropy current divergence. Finally note that the last line of \eqref{System1} is in perfect accord with what we anticipate: the viscosities and conductivities which are the frictional terms, are forced to be zero in the non-dissipative theory.

In \cite{Jensen:2011xb} a simple interpretation of the functions ${\mathfrak M}_\Omega$ and ${\mathfrak M}_B$  was given: one considers a generalized Gibbs ensemble where the magnetic field $B$ and vorticity $\Omega$ are viewed as thermodynamic parameters. In such an ensemble the functions $\{ {\mathfrak M}_B, {\mathfrak M}_\Omega\}$ are defined to be the conjugate variables to $\{B,\Omega\}$ respectively. To wit, assuming a generalized Gibbs-Duhem relation for the thermodynamic pressure $P(T,\mu,B,\Omega)$ one defines the new functions:
\begin{align}
  dP &= s\, dT + \rho \, d\mu -{\mathfrak M}_B \, dB- {\mathfrak M}_\Omega \, d\Omega \notag \\
   \Rightarrow  &\;\;
 {\mathfrak M}_\Omega = -\frac{\partial P}{\partial \Omega} \, ,\qquad
{\mathfrak M}_B = -\frac{\partial P}{\partial B} \, . \label{PhysInterp}
\end{align}
We will return to this when we compare these results against those derived using the effective action.

To summarize, this analysis we began with an arbitrary parity-odd fluid, which is parameterized by $11$ transport coefficients ($5$ in energy-momentum and $6$ in the current) along with $9$ parameters in the entropy current (including the conserved pseudo-vector described in \eqref{DivFreeVector}). Demanding that we have a divergence free entropy current forces us onto a $4$-parameter family of transport (plus an extra parameter in the entropy current multiplying the divergence-free vector): two of these are the functions ${\mathfrak M}_B(T,\bar\mu)$ and ${\mathfrak M}_\Omega(T,\bar\mu)$ which we described above and the remaining two are the Hall viscosity and Hall conductivity ${\tilde \eta}$ and ${\tilde \sigma}$ respectively, which do not enter the entropy analysis. All the parity-odd transport data ${\tilde \chi}_T$, ${\tilde \chi}_E$, ${\tilde \chi}_\Omega$, ${\tilde \chi}_B$ are determined by the two functions ${\mathfrak M}_B(T,\bar\mu)$ and ${\mathfrak M}_\Omega(T,\bar\mu)$
we introduced. This is qualitatively similar to the result obtained in \cite{Jensen:2011xb}
by demanding the more physical requirement of the second law being locally satisfied. The main difference of course is that for non-dissipative fluids the parity-even transport coefficients $\{\zeta, \eta, \chi_T, \chi_E, \sigma\}$ are forced to vanish.

%~~~~~~~~~~~~~~~~~~~~~~~~~~~~~~~~~~~~~~~~~~~~~~~~~~~~~~~~~~
\subsection{Comparison of the two approaches}
\label{sec:3compare}
%~~~~~~~~~~~~~~~~~~~~~~~~~~~~~~~~~~~~~~~~~~~~~~~~~~~~~~~~~~

We can now compare the results from the effective action analysis of parity-odd charged fluids \S\ref{sec:3eaodd} to the results from the non-dissipativeness constraint in the entropy current formalism \S\ref{sec:3ecodd}.

The effective action formalism parameterizes parity-odd $2+1$ dimensional fluids in terms of two free functions ${\mathfrak w}(s,\mu)$ and ${\mathfrak b}(s,\mu)$.  The entropy current analysis on the other hand led to four free functions which determine transport, viz., $\{{\mathfrak M}_\Omega, {\mathfrak M}_B , {\tilde \eta}, {\tilde \sigma} \}$. We already thus see that the effective action constrains dynamics more strongly than the existence of a conserved entropy current.
 
To understand the mismatch let us explore how the set of parameters $\{{\mathfrak w}, {\mathfrak b} \}$ are related to $\{{\mathfrak M}_\Omega, {\mathfrak M}_B , {\tilde \eta}, {\tilde \sigma} \}$.
Doing so however is not completely straightforward, since we have to account for the difference in the fluid frames in which these results are derived. The effective action result is obtained naturally in the entropy frame whereas the entropy current analysis
has been done in Landau frame. 

In order to compare the results, we do not have to perform the frame change explicitly. Instead we can extract the frame invariant content of the expressions (\ref{Pi1}, \ref{Current1}) and
(\ref{TEntr}, \ref{JEntr}), respectively. 
The frame invariant scalar, vector and tensor data for first order stress tensor and current corrections
are well known \cite{Bhattacharya:2011tra}:\footnote{In these formulas the factors of $\frac{1}{2}$ should be replaced by $\frac{1}{d-1}$ in higher dimensions.}
\begin{align}
   C_S &= \frac{1}{2} P_{\alpha\beta} \Pi^{\alpha\beta} - \left[\frac{\partial P}{\partial \varepsilon}\right]_q  \, u_\alpha u_\beta\, \Pi^{\alpha\beta}
          + \left[\frac{\partial P}{\partial q}\right]_\varepsilon u_\alpha \nu^\alpha \,,\\
   C_V^\alpha &= P^{\alpha}_\beta \left( \frac{q}{\varepsilon+P} \, u_\rho \Pi^{\rho\beta} + \nu^\beta \right)\,,\\
   C_T^{\alpha\beta} &= P^\alpha{}_\rho P^\beta{}_\sigma \Pi^{\rho\sigma} - \frac{1}{2} P^{\alpha\beta} 
          P_{\rho\sigma} \Pi^{\rho\sigma} \, .
\end{align}

Equating the respective frame invariant data from the two different approaches, we can determine the following transport coefficients 
from the general entropy current in terms of the two functions in the effective action (see appendix \ref{appendix:calc} for details):
\begin{align}
\tilde \chi_\Omega &= \left[\frac{\partial P}{\partial \varepsilon}\right]_q \left( 2\tilde{{\mathfrak w}} - T \frac{\partial \tilde{{\mathfrak w}}}{\partial T}
       - \mu \frac{\partial \tilde{{\mathfrak w}}}{\partial \mu}\right) + 
       \left[\frac{\partial P}{\partial q}\right]_\varepsilon \left(\tilde{{\mathfrak b}}- \frac{\partial \tilde{{\mathfrak w}}}{\partial \mu} \right) 
          \, ,\notag\\
\tilde \chi_B &= \left[\frac{\partial P}{\partial \varepsilon}\right]_q \left(\tilde{{\mathfrak b}}-T \frac{\partial \tilde{{\mathfrak b}}}{\partial T}
         - \mu \frac{\partial \tilde{{\mathfrak b}}}{\partial \mu}\right) - \left[\frac{\partial P}{\partial q}\right]_\varepsilon 
         \frac{\partial \tilde{{\mathfrak b}}}{\partial \mu} \, ,\notag\\
T \tilde \chi_T &= \left(\tilde{{\mathfrak b}} - T \frac{\partial \tilde{{\mathfrak b}}}{\partial T} - \mu \frac{\partial \tilde{{\mathfrak b}}}{\partial \mu} \right)
       - \frac{q}{\varepsilon + P} \left( 2\tilde{{\mathfrak w}} - T \frac{\partial\tilde{{\mathfrak w}}}{\partial T} - \mu \frac{\partial\tilde{{\mathfrak w}}}{\partial \mu} \right)  \, ,\notag \\
\tilde \chi_E &= -\frac{\partial \tilde{{\mathfrak b}}}{\partial \mu} +\frac{q}{\varepsilon +P} \left( \frac{\partial \tilde{{\mathfrak w}}}{\partial\mu} -\tilde{{\mathfrak b}}\right) \,,\notag\\
\tilde \sigma+ \tilde \chi_E  &= -\frac{2q}{\varepsilon +P} \left( \tilde{{\mathfrak b}} - \frac{q}{\varepsilon +P} \, \tilde{{\mathfrak w}}\right)\, ,\notag\\
\chi_E &= \chi_T = \sigma = \eta =\zeta = \tilde \eta = 0 \, , \label{System2}
\end{align}
where derivatives with respect to $\mu$ (or $T$) are now taken at constant $T$ (or $\mu$), i.e.\ we treat 
the $s$-dependence of ${\mathfrak b}(s,\mu)$ and ${\mathfrak w}(s,\mu)$ as a dependence on the conjugate variable $T$: $s=s(T,\mu)$. This means we implicitly perform a Legendre transformation at the level of the Lagrangian and denote the Legendre-transformed parameters as $\tilde{{\mathfrak b}}(T,\mu)$ and $\tilde{{\mathfrak w}}(T,\mu)$.
%\begin{equation}\label{eq:legendre}
% \tilde{S}_1 \equiv \int d^3x\, \sqrt{-g} \, \left[ {\mathfrak w}(s,\mu) \, \Omega +{\mathfrak b}(s,\mu) \, B + s\, T\right]
% =\int d^3x\, \sqrt{-g} \, \left[ \tilde{{\mathfrak w}}(T,\mu) \, \Omega +\tilde{{\mathfrak b}}(T,\mu) \, B \right]\,,
%\end{equation}
%where $T \equiv - \frac{\delta}{\delta s} S_1$.
Note that the expressions (\ref{System2}) closely resemble the form of equations (1.8) in \cite{Jensen:2011xb}.

We note that while $\tilde \sigma$ and $\tilde \eta$ were left undetermined by the condition of vanishing 
entropy production, they do get constrained in the effective action approach. 
In fact, the full systems of equations (\ref{System1}) and (\ref{System2}) are completely compatible.
This is achieved by expressing the free parameters of the entropy current analysis 
(i.e., the functions $\mathfrak{M}_B$, $\mathfrak{M}_\Omega$
and the two undetermined transport coefficients $\tilde \sigma$, $\tilde \eta$) in terms
of the two functions in the effective action:
\begin{equation}
  \begin{aligned}
  \mathfrak{M}_\Omega &= -\tilde{{\mathfrak w}} \,,\qquad
  \mathfrak{M}_B = -\tilde{{\mathfrak b}} \, ,\qquad 
  \tilde \eta = 0\,, \\
  \tilde \sigma+\tilde \chi_E[\tilde{{\mathfrak b}},\tilde{{\mathfrak w}}] &=  - \frac{2\,q}{\varepsilon +P} \left(\tilde{{\mathfrak b}} - \frac{q}{\varepsilon +P} \, \tilde{{\mathfrak w}}\right) \, .
  \end{aligned}
  \label{Matching} 
\end{equation}
This result can easily be understood equivalently in terms of the physical interpretation of $\mathfrak{M}_B$ and $\mathfrak{M}_\Omega$  in eq.\ (\ref{PhysInterp}): In an equilibrium configuration, the Euclidean effective action on a time circle of circumference 
$\tfrac{1}{T}$ can be considered as the thermodynamic potential which corresponds to 
(minus) free energy. By working in a stationary background \cite{Banerjee:2012iz, Jensen:2012jh} describe an effective action (functional of background sources) that captures the physics of adiabatic transport. In this formalism, the equilibrium partition function of first order, parity-odd charged fluids in $2+1$ dimensions is described by two functions of background sources (denoted $\alpha$, $\beta$ in \cite{Banerjee:2012iz}; cf., their Eq (1.16)). These are simply related to our parameterization in terms of $\tilde{{\mathfrak w}}, \tilde{{\mathfrak b}}$. Comparing the formalisms it is easy to check that $\alpha\, T = -\tilde{{\mathfrak b}}$ and $\beta\, T^2 = -\mu\, \tilde{{\mathfrak b}}+ \tilde{{\mathfrak w}}$. In effect the thermodynamic functions $\tilde{{\mathfrak w}}$ and $\tilde{{\mathfrak b}}$ can be viewed as the response of the fluid to background vorticity and magnetic field in a linear response sense. Effectively one can write the Euclidean action on the thermal circle to first order in gradients as:
\begin{align}
S_\text{(1)}^\text{(Eucl.)} &= -(\varepsilon - sT) = -  \big( q\mu - P(T,\mu,B,\Omega)\big) \\
 \Rightarrow \quad dS_\text{(1)}^\text{(Eucl.)} &= s \, dT - \mu \, dq - 
   \mathfrak{M}_B \, dB - \mathfrak{M}_\Omega\, d\Omega \,.
\end{align}
and use the fact that the derivatives of the (Legendre transform of) effective action (\ref{1stOrderAction}) with respect to $B$ and $\Omega$ give just the functions $\tilde{{\mathfrak b}}$ and $\tilde{{\mathfrak w}}$, respectively. Note that we are working to leading order in the magnetic field and vorticity and as a result will not be able to see the classical contribution to the response obtained from demanding steady state behaviour as described in \cite{Leigh:2012jv}.

This result suggests that the existence of a fluid action in the naturally non-dissipative effective action formalism provides stronger constraints on hydrodynamics than those obtained from an entropy current analysis. 

%~~~~~~~~~~~~~~~~~~~~~~~~~~~~~~~~~~~~~~~~~~~~~~~
\section{Discussion}
\label{sec:discuss}
%~~~~~~~~~~~~~~~~~~~~~~~~~~~~~~~~~~~~~~~~~~~~~~

We have examined the constraints on hydrodynamics by examining the relatively simple case of charged, non-dissipative, parity-odd fluids in $2+1$ dimensions. The restriction of non-dissipativity is naturally captured in the effective action approach and is implemented in the standard hydrodynamical entropy current framework, by demanding the presence of a conserved entropy current. 

The effective action approach demonstrates that one has a two parameter family of non-dissipative fluids. These are parameterized by two thermodynamic functions ${\mathfrak b}(s,\mu)$, ${\mathfrak w}(s,\mu)$ or their Legendre-transforms $\tilde{{\mathfrak b}}(T,\mu)$, $\tilde{{\mathfrak w}}(T,\mu)$. The latter are interpreted as conjugate variables to the background vorticity and magnetic field respectively. On the other hand by a simple generalization of the entropy current analysis of \cite{Jensen:2011xb}, we were able to demonstrate the presence of a {\em four} parameter family of entropy preserving parity-odd fluids. In particular, the entropy current is conserved for any value of the Hall viscosity ${\tilde \eta}$ and Hall conductivity ${\tilde \sigma}$ and allows in addition two more thermodynamic parameters which we parameterized as ${\mathfrak M}_B$ and ${\mathfrak M}_\Omega$ respectively. By carefully comparing the two analyses (equating the frame-independent data) we found the relations \eqref{Matching}. 

As a result we learn that the $2+1$ dimensional, charged, non-dissipative, first order fluids which are captured by the effective action form a subset of those which are consistent with vanishing entropy production: 
For general parameters ${\mathfrak w}(s,\mu)$ and ${\mathfrak b}(s,\mu)$ in the effective action, the four undetermined parameters in the most general non-dissipative entropy current can be determined consistently as described in \eqref{Matching}. The effective action is simply parameterized by the ``conjugate variables'' to background magnetic field $B$ and vorticity $\Omega$, since ${\mathfrak M}_\Omega = -\tilde{{\mathfrak w}}$ and ${\mathfrak M}_B = -\tilde{{\mathfrak b}}$ and all other transport phenomena are determined in terms of the these two thermodynamic functions (and the equation of state). Furthermore, as described above our result is consistent with analysis of equilibrium partition functions and our effective action can be viewed as a generalization of the partition function to include arbitrary time dependence (up to a Legendre transformation). 

We note in passing that one can rewrite our effective action symmetrically in terms of the gauge potential ${\bf A}$ and the transverse {\em shadow hydrodynamic gauge potential} $\hat {\bf A} = {\bf A} + \mu\, {\bf u}$. Consider a rewriting of \eqref{1stOrderAction}
\begin{equation}
S_1^{a} = \int {\mathfrak b}_a \, {\bf u}\wedge d{\bf A} - \int {\hat {\mathfrak b}}_a\, {\bf u}\wedge d{\hat {\bf A}}
\label{}
\end{equation}	
which is easily obtained from \eqref{1stOrderAction} by ${\mathfrak b}\to {\mathfrak b}_a - {\hat {\mathfrak b}}_a$ and ${\mathfrak w} \to -\mu\, {\hat {\mathfrak b} }_a$. The rationale for doing so is that the transverse potential ${\hat {\bf A}}$ makes its appearance in various considerations of anomalous transport in hydrodynamics \cite{Banerjee:2012cr}. 

We now proceed to discuss some of the salient physical features of our analysis, contrasting our results with earlier works on parity-odd transport.

\paragraph{Mystery of the absent Hall viscosity:}
The curious feature of our analysis is that the effective action fails to capture any non-trivial tensor data implying that the relativistic Hall viscosity vanishes ${\tilde \eta} =0$. We do not see an a priori reason for this to have been so. More importantly, there have been earlier suggestions that the effective field theory adapted to neutral fluids provides a model to determine the Hall viscosity \cite{Nicolis:2011ey}. However, the arguments provided in \cite{Nicolis:2011ey} are somewhat indirect; they use an effective action of the form described here with ${\mathfrak w} = {\mathfrak w}(s)$ and ${\mathfrak b} =0$ which is appropriate for a neutral parity-odd fluid. The stress tensor derived from their action is indeed consistent with what we derived in \eqref{Pi1}. One can indeed check that the relativistic stress tensor derived in \cite{Nicolis:2011ey} does not have non-trivial frame invariant tensor data, which it would need to do describe Hall viscosity.\footnote{This result was also consistent with  the analysis reported in  \cite{Bhattacharya:2012zx} where neutral parity-odd fluids were considered (see their Appendix) and the absence of tensor data noted.}

The authors of \cite{Nicolis:2011ey} actually go on to  consider non-relativistic fluids (obtained by taking a suitable scaling limit as described in \cite{Bhattacharyya:2008kq}) and realize that their effective action as such does not contribute to Hall viscosity. They then proceed to modify their constitutive relations with a phenomenologically motivated improvement term. Such a term however does not naturally arise from an action and we are thus forced to conclude that the effective action formalism does not capture the physics of Hall viscosity. 

What exactly does it take for an effective action to capture Hall viscosity? The fact that such a term is admissible in hydrodynamic transport was realized a while ago in the context of quantum Hall systems \cite{Avron:1995fg}. It was called odd-viscosity there and  its contribution to transport was derived by computing the adiabatic Berry curvature picked up as we traverse a degenerate sub-space of the quantum Hilbert space. Its properties were further elaborated upon in \cite{Avron:1997fk} for non-relativistic systems and it has received much interest in the condensed matter literature due to its potential measurability in quantum Hall systems and its contribution to transport have been elaborated on in various works, cf., \cite{Read:2008rn, Haldane:2009ke, Bradlyn:2012ea}.\footnote{See  \cite{Saremi:2011ab, Chen:2011fs,Chen:2012ti} for some holographic computations of Hall viscosity in models where one has axionic coupling to curvature and also \cite{Liu:2012zm} for a more recent discussion.}
 In all these contexts it is clear that the transport associated with the parity-odd Hall viscosity is non-dissipative and indeed this is cleanly borne out from the entropy current analysis \cite{Jensen:2011xb}. Non-negativity of entropy current divergence is insensitive to the value of ${\tilde \eta}$ (to first order in the gradient expansion). So why then is this term not being captured by the effective action?

We note that restricting to equilibrium configurations does not help the matter, since the Hall viscosity term drops out in equilibrium. The easiest way to see this is to note that linear response Kubo formula for Hall viscosity \cite{Saremi:2011ab} (see also \cite{Jensen:2011xb,Bradlyn:2012ea}) illustrates that one has to consider frequency dependence of the stress tensor two point function, which indicates that the Hall viscosity is a transport coefficient and not a thermodynamic response parameter (despite being adiabatic).  One might speculate that Hall viscosity is hard to encode in the effective action despite being adiabatic (i.e., not contributing to entropy production), because of it being an honest transport coefficient. This however cannot be the full story, since as we now explain the anomalous Hall conductivity is indeed captured in the effective action formalism (it is qualitatively similar to the Hall viscosity).

One interesting modification of the effective action to allow for non-vanishing Hall viscosity is to allow for the gravitational background to have non-vanishing torsion. 
This is motivated in part from the earlier observations of \cite{Hughes:2011hv,Hughes:2012vg} and also the more recent analysis of \cite{Hoyos:2013eha}.\footnote{We thank Sergej Moroz for a discussion that inspired this line of thought and R. Loganayagam for useful hints about dealing with torsional connections.} The basic idea would be to allow for the torsion to be an independent degree of freedom and to then ascertain its effects on the energy-momentum and charge transport. 

The idea is that for theories with intrinsic spin, the vielbein and the spin connection should be regarded as independent background sources. For a given affine connection $\Gamma^\lambda{}_{\alpha\beta}$ this is equivalent to treating the metric and the torsion tensor $S^\lambda{}_{\alpha\beta} \equiv \Gamma^\lambda{}_{[\alpha\beta]}$ as independent sources (for a recent review on the necessity of allowing for torsion in order to account for intrinsic spin of matter in gravity see \cite{Poplawski:2013koa}.) The torsional part then gives rise to a new {\it spin contribution} to the stress tensor. While such an approach seems reasonable a priori, we show in Appendix \ref{appendix:torsion} that in our case it does not give rise to any new structures in the total hydrodynamical stress tensor: the new orbital spin contribution cancels out once we define the appropriately symmetrized stress tensor. Note that the spin contribution is generically not symmetric and the stress tensor needs to be improved to obtain a symmetric object. In any event when the dust settles we find that the Hall viscosity is not recovered by this particular extension. 

\paragraph{Vorticity contribution to anomalous Hall conductivity:} Let us turn to another interesting transport coefficient, ${\tilde \sigma}$, the anomalous Hall conductivity. This is so named because it contributes to the transverse current in a fashion similar to the conventional Hall current; however, as we will see momentarily there is a transverse current induced even in the absence of an external magnetic field. Consider the expression for ${\tilde \sigma}$ given in \eqref{Matching}. By writing out the result for ${\tilde \chi}_E$ explicitly, we have 
\begin{equation}
{\tilde \sigma} = -\frac{q}{\varepsilon+P}\, \tilde{{\mathfrak b}} + \frac{1}{T}\, \tilde{{\mathfrak b}}_{,{\bar\mu}}  - \frac{q}{\varepsilon+P}\left(\frac{1}{T}\, \tilde{{\mathfrak w}}_{,\bar \mu} - \frac{2\,q}{\varepsilon+ P}\, \tilde{{\mathfrak w}} \right)
\label{}
\end{equation}	
This in particular shows that there is a contribution to the anomalous Hall conductivity sourced both by the background magnetic field and by the vorticity. Examining the Kubo formula for ${\tilde \sigma}$  \cite{Jensen:2011xb} we again realize that it is an honest transport coefficient. Despite this it is nevertheless captured in the effective action approach. 

\paragraph{Lessons for the effective action:} One of the main motivations for examining the effective action for hydrodynamic transport of non-dissipative fluids is to ascertain whether there are constraints beyond those imposed by the second law of thermodynamics. While we believe this to be the case, we find it hard to argue from the analysis undertaken herein that we have a definitive answer to this question. In part this analysis was carried out to explore the constraints in a more controlled setting than the case of neutral second order hydrodynamic transport described in \cite{Bhattacharya:2012zx}, since one has less intuition for the second order transport coefficients.   It however is puzzling that a physically sensible adiabatic transport data is incapable of being encoded in the effective action. It is tempting to speculate that there is something missing in the effective action approach which overly constrains fluid dynamics; we have some preliminary indication of similar behaviour from study of anomalous transport phenomena \cite{Haehl:2013fk} and hope to elaborate on this issue in the future.

%~~~~~~~~~~~~~~~~~~~~~~~~~~~~~~~~~~~~~~~~~~~~~~
\acknowledgments 
%~~~~~~~~~~~~~~~~~~~~~~~~~~~~~~~~~~~~~~~~~~~~~~
It is a pleasure to thank Jyotirmoy Bhattacharya and Sayantani Bhattacharyya for collaboration during the early stages of the project and for various discussions. It is a pleasure to thank R. Loganayagam for numerous insightful comments and discussions on various aspects of the effective action approach. We would also like to thank Hong Liu for comments on a draft of the paper. 
MR would like to thank CERN, ITF Amsterdam, Crete Center for Theoretical Physics, ICTP, Trieste and Technion, Haifa for hospitality during the course of this project. FH is supported by a Durham Doctoral Fellowship while MR is supported in part by the the STFC Consolidated Grant ST/J000426/1.

\appendix
%~~~~~~~~~~~~~~~~~~~~~~~~~~~~~~~~~~~~~~~~~~~~~~~~~~~~~~~~~~
\section{Frame invariant data from the effective action}
\label{appendix:calc}
%~~~~~~~~~~~~~~~~~~~~~~~~~~~~~~~~~~~~~~~~~~~~~~~~~~~~~~~~~~

This appendix provides some details on the comparison of frame invariant data, eq.\ (\ref{System2}), of the two approaches
to non-dissipative partity-odd hydrodynamics. 

Extracting the frame invariant data from the 
parametrization of the most general first order conserved currents \eqref{TEntr} and \eqref{JEntr} is straightforward to do:\footnote{In writing these expressions we are setting the parity-even contributions to zero as the corresponding transport coefficients vanish in  non-dissipative systems.}
\begin{align}
   C_S  &= \tilde \chi_\Omega \,\Omega + \tilde \chi_B\, B \, , \\
   C_V^\mu &=  \tilde \chi_T \,\tilde V_1^\mu+ (\tilde \chi_E +\tilde \sigma) \,\tilde V_2^\mu - T \, \tilde \sigma \, \tilde V^\mu_3\, , \\
   C_T^{\mu\nu} &= - \tilde \eta\, \tilde \sigma^{\mu\nu} \, ,
\end{align}
where we already set those coefficients to zero which vanish according to the result (\ref{System1}). 

We will now explain how to get the frame invariant data of the effective action result (\ref{Pi1}, \ref{Current1}) into
a useful form. These data are given by 
\begin{align}
  C_S &= \left[ -s\, {\mathfrak w}_{,s} - \left[ \frac{\partial P}{\partial \varepsilon}\right]_q \left(\mu \, {\mathfrak w}_{,\mu}-2{\mathfrak w}\right)
    + \left[ \frac{\partial P}{\partial q}\right]_\varepsilon \left( {\mathfrak b}- {\mathfrak w}_{,\mu}\right) \right] \Omega  \notag\\
      &\quad+ \left[ - s\, {\mathfrak b}_{,s} - \left[ \frac{\partial P}{\partial \varepsilon} \right]_q \left(\mu\, {\mathfrak b}_{,\mu} - {\mathfrak b}\right) 
      - \left[ \frac{\partial P}{\partial q}\right]_\varepsilon {\mathfrak b}_{,\mu} \right] B \; , \label{CScalar}\\
  C_V^\alpha &= \Bigg[ \frac{1}{T} {\mathfrak b} - \frac{\mu}{T} {\mathfrak b}_{,\mu} - \left( \left[\frac{\partial s}{\partial T}\right]_\mu +
     \frac{\mu}{T} \left[\frac{\partial s}{\partial \mu}\right]_T\right) {\mathfrak b}_{,s} \notag \\
     & \qquad + \frac{q}{\varepsilon + P} \left( -\frac{2}{T} {\mathfrak w} + \frac{\mu}{T} {\mathfrak w}_{,\mu}
     + \left( \left[\frac{\partial s}{\partial T}\right]_\mu + \frac{\mu}{T} \frac{\partial s}{\partial \mu} \right) {\mathfrak w}_{,s} \right)\Bigg] \tilde V_1^\alpha\notag\\
     &\quad+ \left[ 2\left( \frac{q}{\varepsilon +P}\right)^2 {\mathfrak w} - \frac{2q}{\varepsilon +P} {\mathfrak b}\right] \tilde V_2^\alpha \notag\\
     &\quad- \left[{\mathfrak b}_{,\mu} + \frac{\partial s}{\partial \mu} {\mathfrak b}_{,s} - \frac{q}{\varepsilon +P} \left({\mathfrak b}+{\mathfrak w}_{,\mu} + \frac{\partial s}{\partial \mu} 
     {\mathfrak w}_{,s} \right) + 2 \left( \frac{q}{\varepsilon +P} \right)^2 {\mathfrak w} \right] T\, \tilde V_3^\alpha \; ,\\
 C_T^{\alpha\beta} &= 0 \; ,
\end{align}
where, in order to rewrite the vector piece in the basis $(\tilde V_1^\alpha, \, \tilde V_2^\alpha,\, \tilde V_3^\alpha)$, we used 
the off-shell identities 
\begin{align}
  \eps^{\alpha\rho\sigma} \nabla_\rho u_\sigma &= - \eps^{\alpha\rho\sigma}\,
  u_\rho \,{\mathfrak a}_\sigma - \Omega\, u^\alpha \, , \\
  \eps^{\alpha\rho\sigma} \nabla_\rho A_\sigma &= \eps^{\alpha\rho\sigma} \,u_\rho \,E_\sigma - B\, u^\alpha \,, 
\end{align}
as well as the following formulas which already implement a partial Legendre transformation from an entropic description in terms of $s$ to a description in terms of the conjugate variable $T$ by taking $s=s(T,\mu)$:
\begin{align}
   \eps^{\alpha\rho\sigma}\,
  u_\rho \, {\mathfrak a}_\sigma &=-\frac{1}{T}\tilde V_1^\alpha + \frac{q}{\varepsilon +P}\left( \tilde V_2^\alpha - T \, \tilde V_3^\alpha\right)
   \, , \label{identS}\\
  \nabla_\rho s &= s\,\Theta \, u_\rho + \frac{1}{s} \left[ \frac{\partial s}{\partial T}\right]_\mu \left[
     q \,(E_\rho -P^\sigma_\rho \,\nabla_\sigma \mu)- (\varepsilon +P) \, {\mathfrak a}_\rho \right]
     + \left[ \frac{\partial s}{\partial \mu} \right]_T P^\sigma_\rho \,\nabla_\sigma \mu \, .\label{identS2}
\end{align}
Equations (\ref{identS}, \ref{identS2}) can be derived using the zeroth order fluid equation and the first law in the form $dP = s \,dT + q \, d\mu$. 
Indeed, the perfect fluid stress tensor conservation reads
\begin{align} \label{PerfFluid}
\nabla_\beta T_{(0)}^{\alpha\beta}=F^\alpha{}_\beta J^\beta_{(0)} 
 \qquad &\Rightarrow\qquad (\varepsilon+P) \, {\mathfrak a}^\alpha + s \, P^{\alpha\beta} \nabla_\beta T + q \,P^{\alpha\beta} \nabla_\beta \mu =q\,E^\alpha\, .
\end{align}
Multiplying this equation with $\eps_{\alpha\beta\gamma}u^\beta$, one directly obtains (\ref{identS}). Eq.\ (\ref{identS2}) has a transverse
part which follows from projecting the conservation equation (\ref{PerfFluid}) with $P^\beta_\alpha$, and a projection onto the fluid velocity which is just the
statement of conservation of the entropy current:
\begin{align}
 \nabla_\alpha (s\, u^\alpha ) = 0 \qquad \Rightarrow \qquad u^\alpha \nabla_\alpha s = - s \, \Theta  \; . 
\end{align}

In order to write the scalar piece (\ref{CScalar}) in the form in which it enters the system (\ref{System2}), we replace the entropy density using 
the following thermodynamic identity which completes the Legendre transformation such that $s=s(T,\mu)$:
\begin{align}
 s = \left[ \frac{\partial P}{\partial T} \right]_\mu =  \left[\frac{\partial P}{\partial \varepsilon}\right]_q \left(
   T \left[\frac{\partial s}{\partial T}\right]_\mu + \mu \left[\frac{\partial s}{\partial \mu}\right]_T \right) + 
   \left[ \frac{\partial P}{\partial q}\right]_\varepsilon \left[ \frac{\partial s}{\partial \mu}\right]_T \; .
\end{align}
%

%~~~~~~~~~~~~~~~~~~~~~~~~~~~~~~~~~~~~~~~~~~~~~~~~~~~~~~~~~~
\section{Hall viscosity \& torsional connections}
\label{appendix:torsion}
%~~~~~~~~~~~~~~~~~~~~~~~~~~~~~~~~~~~~~~~~~~~~~~~~~~~~~~~~~~

In this appendix we show that the two terms in the effective action (\ref{1stOrderAction}) do not give rise to any new stress tensor contributions even when we allow for affine connections whose torsion is non-zero. 

Let us start by reviewing how a connection with torsion leads to changes in the stress-energy tensor. To this end let us
treat the metric and the Christoffel symbols as independent objects and define the \textit{metric} stress tensor $t^{\alpha\beta}$ 
from variations with respect to the metric alone:
\begin{align}
 \delta S = \int \sqrt{-g} \, \left( J^\alpha\, \delta A_\alpha  + \frac{1}{2} t^{\alpha\beta}\, \delta g_{\alpha\beta} + 
 \frac{1}{2} X^{\lambda \alpha}{}_\beta \, \delta \Gamma^\beta{}_{\lambda\alpha} \right) \,,
\end{align}
where $\Gamma^\beta{}_{\lambda\alpha}$ defines some arbitrary affine connection (in particular, $\Gamma^\beta{}_{\lambda\alpha}$ does not have to
be symmetric in its lower indices). If $\Gamma^\beta{}_{\lambda\alpha}$ were the standard Christoffel symbols associated with the Levi-Civita connection, then 
an integration by parts would yield 
\begin{align}
 \delta S = \int \sqrt{-g}\, \left( J^\alpha\,\delta A_\alpha + \frac{1}{2} T^{\alpha\beta} \, \delta g_{\alpha\beta} \right) \,
\end{align}
with
\begin{align}
 T^{\alpha\beta} = T^{(\alpha\beta)}= t^{\alpha\beta} - \frac{1}{2} \nabla_\lambda X^{\lambda\beta\alpha} +\frac{1}{2} \nabla_\lambda \left(
   X^{\alpha[\beta\lambda]} + X^{\beta[\alpha\lambda]} + X^{\lambda[\beta\alpha]} \right) \,. \label{eq:Tsymm}
\end{align}
It has, however, been motivated that the expression (\ref{eq:Tsymm}) is the correct total canonical 
stress tensor even if the connection is not the Levi-Civita one, see e.g.\ \cite{Hehl:1976kj}. The second term on the right hand side of 
(\ref{eq:Tsymm}) corresponds to an asymmetric orbital contribution to the stress tensor in spinor field theories. 
The three terms in the bracket should be seen as an additional piece 
that comes from a Belinfante-Rosenfeld symmtrization. This amounts to saying that
the terms in the bracket provide a symmetrization of the stress tensor which is consistent
with a pseudo-gauge ambiguity in the definition of the canonical stress tensor from Noether's 
argument \cite{Hehl:1976vr}. This symmetrization improved stress tensor is thus the \textit{canonical} one from Noether's argument
which we should use to make a comparison with the entropy current analysis.

In our discussion of the effective action (\ref{1stOrderAction}) in the main text, we were assuming that the Christoffel symbols are
given by their standard expressions, so they could be completely ignored (their lower indices
were always contracted with a totally antisymmetric tensor).
We can now ask whether allowing for asymmetric Christoffel symbols in 
the effective action (\ref{1stOrderAction}) leads to new terms in the stress tensor as defined by eq.\ (\ref{eq:Tsymm}). 
However, it is easy to see that the $X^{\lambda\alpha}{}_\beta$ as they would occur in 
the action (\ref{1stOrderAction}) would be antisymmetric in their first two indices. 
In that case eq.\ (\ref{eq:Tsymm}) just collapses to $T^{\alpha\beta} = t^{\alpha\beta}$.
We conclude that even for connections with torsion our effective action does not
give spin contributions to the stress tensor which would be needed to account for Hall viscosity.

 %%%%%%%%%%%%%%%%%%%%%%%%%%%%%%%%%%%%%%%%%%%%%%%%

% \bibliographystyle{JHEP}
% \bibliography{non-dissipative}

\providecommand{\href}[2]{#2}\begingroup\raggedright\endgroup
\end{document}